# POWER SPECTRUM CONSTRAINTS FROM SPECTRAL DISTORTIONS IN THE COSMIC MICROWAVE BACKGROUND [†]

Wayne Hu, Douglas Scott, and Joseph Silk

*Departments of Astronomy and Physics, and*
*Center for Particle Astrophysics,*
*University of California, Berkeley, California 94720*

Using recent experimental limits on $\mu$ distortions from COBE FIRAS, and the large lever-arm spanning the damping of sub-Jeans scale fluctuations to the scale of the COBE DMR fluctuations, we set a constraint on the slope of the primordial power spectrum $n$. It is possible to analytically calculate the contribution over the full range of scales and redshifts, correctly taking into account fluctuation growth and damping as well as thermalization processes. We find that the 95% upper limit is weakly dependent on cosmological parameters, *e.g.*, $n < 1.54$ ($h = 0.5$) and $n < 1.56$ ($h = 1.0$) for $\Omega_0 = 1$ with marginally weaker constraints for $\Omega_0 < 1$ in a flat $\Omega_0 + \Omega_\Lambda = 1$ universe.

*Subject Headings:* Cosmology: Cosmic Microwave Background, Cosmology: Early Universe, Cosmology: Theory

hu@pac2.berkeley.edu

# 1. Introduction

Two years of data from the COBE DMR experiment has yielded a firm large scale normalization of the power spectrum and indications that the spectral index may be $n > 1$. The temperature fluctuations on a $10°$ scale have been measured to be $(\Delta T/T)_{rms}(10°) = 1.12 \pm 0.10 \times 10^{-5}$, and the best fit spectral index is $n = 1.42 \pm 0.35$ (Bennett et al. 1994, Wright et al. 1994b). The Tenerife experiment (Hancock et al. 1994) measures a high fluctuation level on scales of $5°$ that, together with the first year COBE DMR result, already requires $n \gtrsim 0.9$. Furthermore, the COBE FIRAS experiment has stringently constrained spectral distortions of the chemical potential type to be $|\mu| < 3.3 \times 10^{-4}$ (95% CL, Mather et al. 1994). We shall show below that, regardless of any other constraints from large-scale structure observations on $n$, a value of $n \gtrsim 1.5$ results in excessive $\mu$–distortions of the spectrum for adiabatic perturbations.

On small scales, photon diffusion dissipates fluctuations and thereby creates spectral distortions (Sunyaev & Zel'dovich 1970; Daly 1991; Barrow & Coles 1991). Daly (1991) noted that diffusion effectively superimposes blackbodies from regions with different temperatures, resulting in a distortion initially of the Compton-$y$ type with amplitude approximately the square of the temperature dispersion (Zel'dovich, Illarionov, & Sunyaev 1972). Wright et al. (1994a) applied this method to the recent COBE results, but did not give a proper treatment of the matter and radiation transfer functions, which relate the fluctuations at any given time to the primordial power spectrum.

Compton scattering processes the initial distortion into a $\mu$ distortion in a manner which depends *sensitively* on the balance between the perturbed energy and number density of photons. Because both quantities are of order the variance of the temperature distribution, the calculation is effectively second order (Hu, Scott, & Silk 1994). The superposition function must take into account complications such as the energy associated with the coherent motions and pressure gradients of the fluctuations. Since the behavior of $\mu$ distortions is independent of the detailed form for the initial spectral perturbation, we instead base our treatment on an accurate account of the number and energy density alone (Sunyaev & Zel'dovich 1970).

In this Letter, we improve on the results of Daly (1991) and Barrow & Coles (1991) by explicitly including the effect of thermalization and the damping of fluctuations from all wavelengths at all times; prior treatments considered the effect to arise from a single wavelength instantaneous perturbation to the spectrum at the thermalization epoch. Furthermore, we normalize the power spectrum to the COBE DMR detection at $10°$. Our treatment uses the correct transfer functions both at small scales, to determine the physics of damping, and at large scales, to fix the normalization. We find that the resulting constraint on the spectral index $n$ of the matter power spectrum $P(k) = Ak^n$ is significantly more stringent than previously inferred.

# 2. Spectral Distortions

Before redshift $z_y = 2.15 \times 10^4 \Theta_{2.7}^{1/2} (\Omega_B h^2)^{-1/2}$ (Burigana et al. 1991), Compton scattering rapidly establishes equilibrium with the electrons, following an injection of energy or creation of photons. Here $h = H_0/100 \mathrm{~km~s^{-1}~Mpc^{-1}}$, $\Theta_{2.7} = T_0/2.7K$, and the present temperature of the CMB is $T_0 = 2.726 \pm 0.005 K$ (Mather et al. 1994). Equilibrium requires that the photon distribution function $f$ is given by a Bose-Einstein spectrum, $f = [\exp(\mu + h\nu/kT_e)]^{-1}$, where $T_e$ is the temperature of the electrons. After $z_y$, equilibrium cannot be established, and any further energy injection will cause a spectral distortion which is approximately of the Compton-$y$ type. True thermal equilibrium with $f$ as a Planck spectrum ($\mu = 0$) cannot in general be attained since Compton scattering does not change the number of photons in the spectrum. However, higher order processes such as bremsstrahlung ($e^- + X \rightarrow e^- + X + \gamma$, $X$ is any ion) and double Compton scattering ($e^- + \gamma \rightarrow e^- + \gamma + \gamma$) can create photons and establish thermal equilibrium. For a universe with $\Omega_B h^2 \lesssim 0.1$ as required by nucleosynthesis, double Compton scattering dominates the thermalization process (Danese & de Zotti 1982, Hu & Silk 1993).

We can therefore follow the creation and evolution of spectral distortions by balancing the energy injection caused by dissipation with the photon creation caused by double Compton scattering. The computational technique, pioneered by Sunyaev & Zel'dovich and refined by Danese & de Zotti (1977), involves solving the evolution equations for the photon number and energy density. Assuming that photon creation is purely due to double Compton scattering, we obtain (Hu & Silk 1993),

$$\frac{d\mu}{dt} \approx -\frac{\mu}{t_{DC}} + 1.4\frac{Q}{\rho_\gamma}, \qquad (1)$$

where $t_{DC} = 2.09 \times 10^{33}(1 - Y_p/2)^{-1}(\Omega_B h^2)^{-1}\Theta_{2.7}^{-3/2}z^{-9/2}$ s is the double Compton thermalization time, $Y_p \approx 0.23$ is the primordial mass fraction in helium, and $Q/\rho_\gamma$ is the fractional rate of energy injection. Equation (1) has the simple solution

$$\mu \approx 1.4 \int_0^{t(z_y)} dt \frac{Q(t)}{\rho_\gamma} \exp[-(z/z_\mu)^{5/2}], \qquad (2)$$

where $z_\mu = 4.09 \times 10^5(1 - Y_p/2)^{-2/5}\Theta_{2.7}^{1/5}(\Omega_B h^2)^{-2/5}$ is the thermalization redshift for double Compton scattering.

### 3. Damping of Small–Scale Fluctuations

The damping of primordial density perturbations before recombination injects energy into the CMB spectrum and leads to spectral distortions. Furthermore, given the current stringent limit on spectral distortions, the arguments of Daly (1991) show that photon diffusion as opposed to nonlinear dissipation is the dominant mechanism for the creation of distortions.

Let us now examine the photon diffusion mechanism. At early times, photons and baryons are tightly coupled and behave as a single viscous fluid. Any fluctuations in the photon-baryon fluid, whether adiabatic or isocurvature, will oscillate as a sound wave inside the photon-baryon Jeans length. Solving the Boltzmann equation for the photons in the tight coupling limit leads to for the photon energy density perturbation (Peebles & Yu 1970),

$$\Delta(k,t) = \Delta_J(k) \cos(\omega\eta) \exp(-\int d\eta \Gamma), \qquad (3)$$

where $\Delta_J(k)$ is the amplitude of the $k$-mode upon entering the Jeans length, $\eta = \int dt/a$ is the conformal time, $a = 1/(1+z)$ is the scale factor, the dispersion relation for acoustic waves is $\omega = c_s k$, and (Peebles 1980)

$$\Gamma(\eta) = \frac{k^2}{6 n_e \sigma_T a} \frac{R^2 + 4(R+1)/5}{(R+1)^2}. \qquad (4)$$

Here $R = 3\rho_b/4\rho_\gamma$, $n_e$ is the electron number density, $\sigma_T$ is the Thomson cross section, and $c = 1$. This damping of the acoustic waves will dump energy into the microwave background. The average energy density in a plane acoustic wave is given by $\rho_s \approx \rho_{\gamma b} c_s^2 \langle \delta_{\gamma b}^2 \rangle$, where $\rho_{\gamma b} = \rho_\gamma + \rho_b$ and $\delta_{\gamma b}$ are the density and density perturbation in the photon-baryon fluid, and the brackets denote an average over an oscillation of the acoustic wave. Since $\mu$ distortions arise at $z > z_y > z_{eq}(= 2.50 \times 10^4 \Omega_0 h^2 \Theta_{2.7}^{-4})$, we can take the radiation–dominated limit of these equations, $R = 0$, $c_s^2 = 1/3$, and

$$\langle \delta_{\gamma b}^2 \rangle \approx \langle \Delta^2(k,t) \rangle = \frac{1}{2}\Delta_J^2(k) \exp[-(k/k_D(t))^2], \qquad (5)$$

where $k_D(z) = 2.34 \times 10^{-5}\Theta_{2.7}(1 - Y_p/2)^{1/2}(\Omega_B h^2)^{1/2}z^{3/2}$ Mpc$^{-1}$. Therefore, the rate of fractional energy injection

$$\frac{Q(t)}{\rho_\gamma} = -\sum_k \frac{1}{3}\frac{d\langle \Delta^2(k,t)\rangle}{dt}. \qquad (6)$$

Inserting this expression into equation (2) allows us to calculate the spectral distortion from photon diffusion.

## 4. Constraining the Power Spectrum

Hitherto, our arguments have been fully general and apply to any $\Omega_0$ or spectrum of fluctuations. We will now specialize the argument to the case of primordial adiabatic fluctuations with $P(k) \propto k^n$ in a flat $\Omega_0 + \Omega_\Lambda = 1$ universe. Adiabatic perturbations in the photon-baryon fluid grow as $a^2$ outside the Jeans scale, which is nearly equal to the horizon scale in the radiation–dominated epoch. The time at which a given mode $k$ crosses the horizon is $a_H \propto k^{-1}$. Therefore, in addition to the primordial spectrum, the fluctuations gain a factor $k^{-2}$. This specifies a transfer function such that $\Delta^2(k,t) = T^2(k,t)Ak^n$, which, including photon diffusion within the Jeans scale in the radiation–dominated epoch, is given by (Kodama & Sasaki 1986),

$$T(k,t) = -5a_{eq}k_{eq}^2 k^{-2} \cos(\omega\eta)\exp[-(k/k_D)^2/2], \tag{7}$$

where $k_{eq} = (Ha)_{eq}$ is the comoving wavenumber that crosses the horizon at matter–radiation equality $a_{eq} = 1/(1 + z_{eq})$. The transfer function $T(k,t)$ has been normalized so that at present on large scales, $P(k) = Ak^n$. Therefore for adiabatic fluctuations, $\Delta_J$ is defined as

$$\Delta_J^2(k) = 25 a_{eq}^2 k_{eq}^4 A k^{n-4}. \tag{8}$$

The normalization constant $A$ can be determined by the COBE detection at $10°$.

We now make the further assumption that the COBE DMR detection probes potential fluctuations on the last scattering surface *via* the Sachs-Wolfe effect (Sachs & Wolfe 1967), assumed to arise from scalar perturbations alone. We can then write

$$\left(\frac{\Delta T}{T}\right)^2 = \frac{1}{4\pi}\sum_{\ell=2}^\infty (2\ell+1)W_\ell C_\ell, \tag{9}$$

where the COBE window function is $W_\ell = \exp[-\ell(\ell+1)\sigma^2]$, with $\sigma = 0.0742$ being the gaussian width of the $10°$ FWHM beam and (Efstathiou, Bond & White 1992)

$$C_\ell = \frac{H_0^4\,\Omega_0^{1.54}}{8\sqrt{\pi}\eta_0^{n-1}}\frac{\Gamma\left(\frac{3-n}{2}\right)\Gamma\left(\ell+\frac{n-1}{2}\right)}{\Gamma\left(\frac{4-n}{2}\right)\Gamma\left(\ell+\frac{5-n}{2}\right)}AV_x, \tag{10}$$

where $V_x$ is the volume over which the universe is assumed to be periodic. Equation (10) takes the radiation transfer function at large scales fully into account for an $\Omega_0 = 1$ universe. For $\Omega_0 < 1$ with a cosmological constant, there is a small correction due to the integrated Sachs-Wolfe effect which boosts the low multipoles. Given the present status of the observational uncertainties, its inclusion will not significantly affect our constraints (Efstathiou, Bond, & White 1992).

We are now in a position to evaluate equation (2) explicitly. Both the sum over $k$ and the integral over time can be performed analytically for power–law initial spectra leading to,

$$\mu = 1.4F(n)\frac{V_x}{2\pi^2}\left[k_D^3 \Delta_J^2(k_D)\right]_{z_\mu}, \tag{11}$$

where

$$F(n) = \frac{1}{10}\Gamma[(n+1)/2]\,\Gamma[3(n-1)/5,(z_y/z_\mu)^{5/2}]. \tag{12}$$

If $n$ is significantly greater than unity, the incomplete gamma function $\Gamma(m,x) \to \Gamma(m)$ since $z_y/z_\mu \ll 1$ and $F(n)$ is roughly of order unity.

It is easy to interpret this result. If $n > 1$, the smallest waves carry the most energy, and the distortion comes almost entirely from the waves that damped at the thermalization epoch. Prior to thermalization, no distortion survives due to the rapidity of the double Compton process. In this way, we have not only justified the approximations of Daly (1991) and Barrow & Coles (1991) which involve instantaneous injection from a single $k$-mode, but have rigorously derived the additional $n$-dependent factor $F(n)$.

On the other hand if $n < 1$, the fractional energy injection from dissipation will be a maximum at the latest relevant time, *i.e.* recombination. This implies that the constraint from spectral distortions will come from the upper limit on Compton-$y$ distortions. Nevertheless, this constraint will not be very powerful since the normalization is fixed at large scales. For $n < 1$, the small–scale power decreases with respect to that on large scales, yielding no constraint on a COBE–normalized spectrum.

Note that the dependence on the cosmological parameters $\Omega_0$, $\Omega_B$, $h$ is quite weak: approximately $\mu \propto (\Omega_B^{1/10} h^{6/5})^{1-n} \Omega_0^{(2-n)/2}$. Hence for $n \approx 1$, $\mu$ is completely independent of $h$ and $\Omega_B$. Moreover, $\mu$ is nearly independent of $\Omega_B$ for *all* $n$, since raising $\Omega_B$ makes both the damping length shorter and the thermalization redshift smaller. It is also useful to provide an approximate inversion of equation (11):

$$n \approx 1 + \frac{\ln[K_1 \Omega_0^{-0.46} \mu / (\Delta T/T)_{10°}^2]}{\ln[K_2 (\Omega_B h^2)^{-1/10} (\Omega_0 h^2)^{-1/2} I(\Omega_0)]} \tag{13}$$

where we find the constants $K_1 = 5.6 \times 10^{-3}$ and $K_2 = 8.9 \times 10^5$ and the small logarithmic correction $I(\Omega_0) \approx 1 - 0.085 \ln \Omega_0$. One can verify that this is an excellent approximation within the range $1.0 < n < 2.0$ and the allowable cosmological parameters. Note that the dependence of $n - 1$ on $\mu$ and the normalization is only logarithmic, and its dependence on the cosmological parameters is almost entirely negligible. Even relatively large changes in $\mu$ or the normalization will not greatly affect the constraint on $n$.

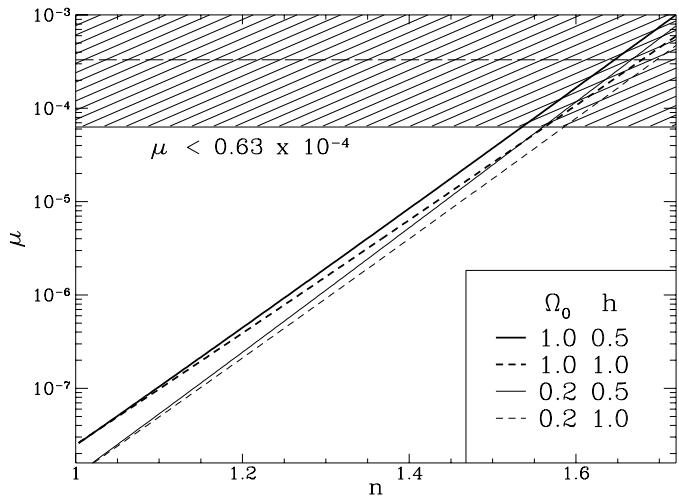

FIG 1. Spectral distortion $\mu$, predicted from the full equation (11), as a function of the power index $n$ for a normalization at the mean of the COBE DMR detection $(\Delta T/T)_{10°} = 1.12 \times 10^{-5}$. With the uncertainties on *both* the DMR and FIRAS measurements, the 95% upper limit is effectively $\mu < 0.63 \times 10^{-4}$ (see text). The corresponding constraint on $n$ is quite weakly dependent on cosmological parameters: $n < 1.54$ ($h = 0.5$) and $n < 1.56$ ($h = 1.0$) for $\Omega_0 = 1$ and quite similar for $0.2 < \Omega_0 = 1 - \Omega_\Lambda < 1$ universes. These limits are nearly independent of $\Omega_B$. We have also plotted (dashed lines) the 95% upper limit on $|\mu| < 3.3 \times 10^{-4}$ for comparison. Note that there is nothing awry with the *placement* of this figure.

The best fit value of $\mu$ to the spectral data from the FIRAS experiment is $\mu = -1.2 \pm 1.1 \times 10^{-4}$ (68% CL) (Mather *et al.* 1994). We can actually make a stronger statement than the quoted FIRAS 95% confidence limit of $|\mu| < 3.3 \times 10^{-4}$, by noting that photon diffusion never creates a negative chemical potential and the current constraint on positive $\mu$ is somewhat tighter than that on $|\mu|$. Taking into account the COBE DMR measurement errors and adopting a $4\mu$K cosmic variance, $(\Delta T/T)_{rms}(10°) = 1.12 \pm 0.18 \times 10^{-5}$ (Bennett *et al.* 1994), we find

$$\frac{\mu}{(\Delta T/T)_{10°}^2} < 5.0 \times 10^5 \quad (95\% \text{CL}). \tag{14}$$

This would be equivalent to an upper limit of $\mu < 6.3 \times 10^{-5}$ for a fixed normalization at the mean value of the DMR detection. Using (14), we set a limit on the slope $n < 1.54$ for $h = 0.5$ and $n < 1.56$ for $h = 1.0$ (see Fig. 1) for $\Omega_0 = 1$ with similar but slightly less stringent limits for cosmological constant–dominated universes ($\Omega_0 < 1$). These constraints are nearly independent of $\Omega_B$ and do not change within the nucleosynthesis bounds of $0.011 < \Omega_B h^2 < 0.016$ (Walker *et al.* 1991, Smith *et al.* 1993).

## 5. Discussion

If the present indications of a large $n$ at COBE DMR scales are borne out by future measurements, we should consider how the $\mu$ constraint might be avoided or at least weakened. If the universe is open ($\Omega_0 + \Omega_\Lambda < 1$), the radiation transfer function *at large scales* (see equation (10)) responsible for the COBE DMR normalization will differ. Significant temperature fluctuations also arise in this model from the integrated Sachs-Wolfe effect (Sugiyama & Gouda 1992, Hu & Sugiyama 1993) which enhances the large scale fluctuations. However, for open universes, there is the more significant effect of an effective cutoff to the temperature perturbations at large scale due to the curvature (Wilson 1983; Kamionkowski & Spergel 1993). This can naturally lead to an apparent $n > 1$ slope at large scales without affecting the slope at small scales. However, note that the dependence of the constraint on the normalization is only logarithmic. In these cases, the upper limit on $n$, the primordial power index, is roughly the same, but does *not* correspond to the effective $n$ that the COBE DMR experiment measures.

To change the limit on the primordial power index $n$, one must change the radiation transfer function *at small scales* where the physics of damping occurs. If the perturbations are not adiabatic, the relation given in equation (7) is not valid and the constraint on $n$ will be altered, *e.g.* for isocurvature fluctuations (Daly 1991). These considerations point to a general way of evading the constraints. We have assumed that the *primordial* power spectrum is a pure power law from the current horizon to the damping scale at thermalization. If we alter the initial conditions so that the effective $n$ decreases sufficiently with increasing $k$, we can always escape constraints from spectral distortions for any given $n$ at large scales.

Aside from these considerations, our constraint is difficult to evade since it relies only on the relatively well understood physics of photon–baryon systems. Similar limits can be applied from other constraints, *e.g.*, primordial black holes (Carr & Lidsey 1993, Novikov *et al.* 1979) or from avoiding excessive anti-bias in galaxies (*e.g.* Lyth & Liddle 1994). However, such limits could be avoided by appealing to our ignorance of the relevant physical processes. By contrast, we believe that the $\mu$ constraint of roughly $n < 1.55$ is a robust limit on pure power laws with adiabatic initial conditions.

This research has been supported at Berkeley in part by grants from NASA and NSF.

FIGURE CAPTIONS

FIG 1. Spectral distortion $\mu$, predicted from the full equation (11), as a function of the power index $n$ for a normalization at the mean of the COBE DMR detection $(\Delta T/T)_{10°} = 1.12 \times 10^{-5}$. With the uncertainties on *both* the DMR and FIRAS measurements, the 95% upper limit is effectively $\mu < 0.63 \times 10^{-4}$ (see text). The corresponding constraint on $n$ is quite weakly dependent on cosmological parameters: $n < 1.54$ ($h = 0.5$) and $n < 1.56$ ($h = 1.0$) for $\Omega_0 = 1$ and quite similar for $0.2 < \Omega_0 = 1 - \Omega_\Lambda < 1$ universes. These limits are nearly independent of $\Omega_B$. We have also plotted (dashed lines) the 95% upper limit on $|\mu| < 3.3 \times 10^{-4}$ for comparison.